\documentclass[runningheads]{llncs}
\usepackage{graphicx}
\usepackage[colorlinks]{hyperref}
\hypersetup{colorlinks, linkcolor=green,citecolor=green}

\begin{document}
\title{Symmetric Dilated Convolution for Surgical Gesture Recognition}

\titlerunning{SD-Conv for Surgical Gesture Recognition}

\author{Jinglu Zhang\inst{1}\and
Yinyu Nie\inst{1} \and
Yao Lyu\inst{1}\and
Hailin Li\inst{2}\and
Jian Chang\inst{1}\and
Xiaosong Yang\inst{1}\and
Jian Jun Zhang\inst{1}
}

\authorrunning{J. Zhang et al.}
%
\institute{National Centre for Computer Animation, Bournemouth University, UK \and
Communication University of China}

\maketitle              
\begin{abstract}
Automatic surgical gesture recognition is a prerequisite of intra-operative computer assistance and objective surgical skill assessment. Prior works either require additional sensors to collect kinematics data or have limitations on capturing temporal information from long and untrimmed surgical videos. To tackle these challenges, we propose a novel temporal convolutional architecture to automatically detect and segment surgical gestures with corresponding boundaries only using RGB videos. We devise our method with a symmetric dilation structure bridged by a self-attention module to encode and decode the long-term temporal patterns and establish the frame-to-frame relationship accordingly. We validate the effectiveness of our approach on a fundamental robotic suturing task from the JIGSAWS dataset. The experiment results demonstrate the ability of our method on capturing long-term frame dependencies, which largely outperform the state-of-the-art methods on the frame-wise accuracy up to $\sim$6 points and the F1@50 score $\sim$6 points.
\end{abstract}
\section{Introduction}
Surgical gesture recognition is the process of jointly segmenting and classifying fine-grained surgical actions from surgical videos. It is crucial for surgical video understanding and building the context awareness system towards the next generation surgery~\cite{maier2017surgical}. However, raw surgical videos are normally untrimmed and the operation environment is particularly complicated. Consequently, detecting surgical gestures from these surgical videos with high intra-class variance and low inter-class variance is inherently quite challenging.

Recent studies apply Hidden Markov Model (HMM)~\cite{tao2013surgical} and its variants~\cite{lea2015improved} to identify the latent state of surgical actions. The latent states transferring among successive actions are subsequently modelled by the transition probability. Although state features in HMMs are interpretable, they only focus on few local frames hence making the model incapable of capturing the global pattern. In addition, some machine learning methods (i.e. Support Vector Machine (SVM)~\cite{twinanda2016endonet}) assemble multiple heterogeneous features (color, motion, intensity gradients, etc.) to localize and classify surgical actions. Nonetheless, these features are hand-crafted. Therefore some crucial latent features could be neglected during feature extraction procedure.

More recently, large number of approaches depend on Recurrent Neural Networks (RNN)~\cite{singh2016multi,dipietro2016recognizing}, particularly, the Long Short Term Memory (LSTM) network, because of their notable ability of modeling sequence data in variable length. The gate mechanism of LSTM preserves temporal dependencies and drops irrelevant information during the training stage. However, LSTM-based methods only have limited ability of capturing long-term video context, due to the intrinsic vanishing gradient problem~\cite{pascanu2013difficulty}.

From another perspective, Lea et al.~\cite{lea2017temporal} introduce Temporal Convolutional Networks (TCNs) to segment and detect actions by hierarchically convolving, pooling, and upsampling input spatial features using 1-D convolutions and deconvolutions. The promising experiment results manifest that TCNs are capable of dealing with long-term temporal sequences though, the model handles information among local neighbors, thus showing incapabilities in catching global dependencies. Following this work, Farha and Gall~\cite{farha2019ms} suggest a multi-stage TCN, in which each stage is composed of several dilation layers, for action segmentation. Their work demonstrates the competence of dilated convolution~\cite{oord2016wavenet} in hierarchically collecting multi-scale temporal information without losing dimensions of data. Moreover, in order to sequentially capture the video dynamics, Funke et al.~\cite{funke2019using} randomly sample video snippets (16 consecutive frames per snippet) and utilize a 3D Convolutional Neural Network (CNN) to extract the spatial-temporal features. But still, they only consider local continuous information. Because of the huge computational cost and GPU memory expenditure of 3D-CNN, they can only train the network at the clip level rather than inputted with the whole video~\cite{zhang2020v4d}.

In this paper, we propose a symmetric dilated convolution structure embedded with self-attention kernel to jointly detect and segment fine-grained surgical gestures. Figure~\ref{figure_1} is an overview of our framework. Taking the extracted spatial CNN features from~\cite{lea2017temporal} as input, the encoder captures the long temporal information with a series of 1-D dilated convolutions to enlarge the temporal receptive field, followed by an attention block to establish the one-to-one relationship across all latent representations. Symmetrically, we devise our decoder with another set of dilation layers to map the latent representations back to each frame and predict the frame-wise gesture label. Unlike 3D-CNN learning features from partial sampled clips, our network takes the whole video into consideration. Owing to the symmetric dilated convolution structure with the enclosed self-attention kernel, not only can we learn the long-range temporal information, but also we can process neighbor and global relationship simultaneously.

With the above facts, we claim our contribution as two-fold. First, we propose a symmetric dilation architecture embedded with a self-attention module. It takes into account the long-term temporal patterns and builds frame-to-frame adjacent as well as global dependencies from the surgical video sequence. Second, with the novel network architecture, our approach consistently exceeds the state-of-the-art method both on frame-level and on segmental-level metrics, improving the frame-wise accuracy \textbf{$\sim$6 points}, and the F1@50 score \textbf{$\sim$6 points}, which largely alleviates the over-segmentation error.

\begin{figure}[htb!]
\centering
\includegraphics[width=1.0\textwidth]{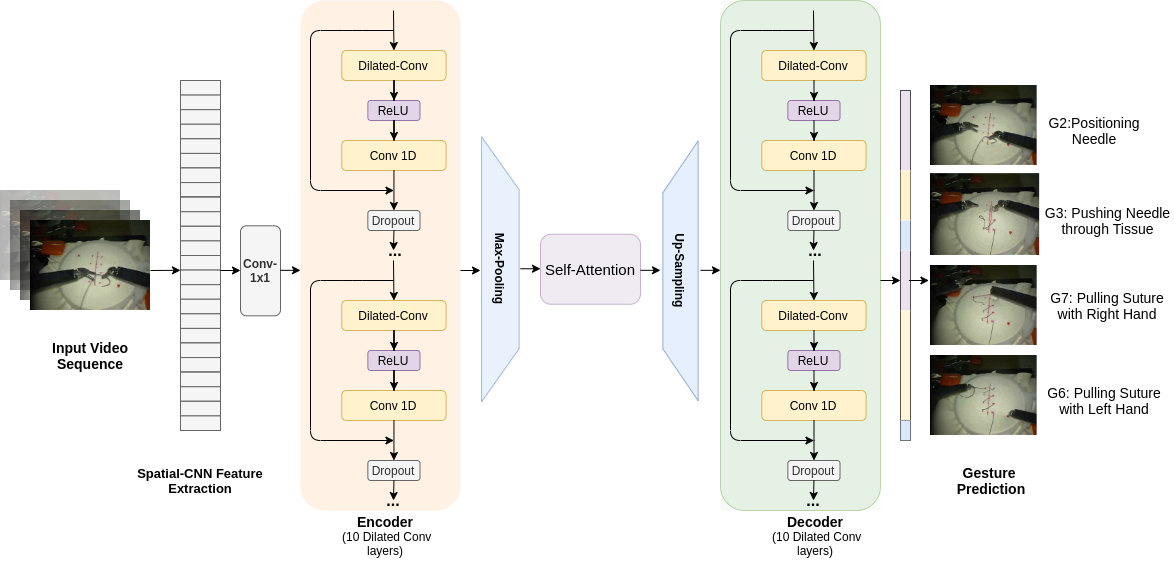}
\caption{Overview of our architecture. Symmetric dilation network takes frame-level spatial-CNN features as input. The architecture can be divided into five steps: 1) 1-D convolution; 2) 10 dilated convolution layers with max-pooling; 3) self-attention; 4) upsampling with 10 dilated convolution layers; 5) frame-wise prediction.}
\label{figure_1}
\end{figure}

\section{Methodology}
The architecture of our symmetric dilation network for surgical gesture recognition is detailed in this section (see Figure~\ref{figure_2}), which consists of two substructures: 1) the symmetric dilated Encoder-Decoder structure to capture long-term frame contents with memory-efficient connections (dilated layers) to aggregate multi-scale temporal information (see section~\ref{STDC}); 2) the self-attention kernel in the middle to deploy the deep frame-to-frame relations to better discriminate the similarities among different frames (see section~\ref{SA}).

\subsection{Symmetric Temporal Dilated Convolution}\label{STDC}
\begin{figure}[htb!]
\centering
\includegraphics[width=0.6\textwidth]{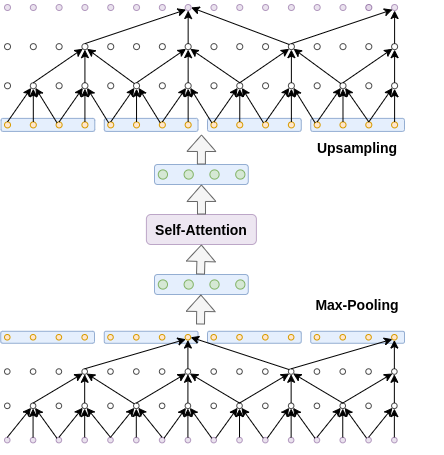}
\caption{Symmetric temporal dilated convolution. With the layer number increasing, the size of the temporal receptive field grows exponentially.}
\label{figure_2}
\end{figure}
Temporal dilated convolution is a type of convolution applied on the input sequence with a defined sliding gap, which increases the temporal receptive field with less parameters. The first layer of the encoder is a $1\times 1$ convolution to map the dimension of input spatial-CNN features to number of kernels $f$, followed by $l$ layers of temporal dilated convolutions, where the dilation rates $\{s_{l}\}$ are set to $s_{l} = 2^l, l=0,1,...,9$. Because our target is the off-line recognition, we follow the details in~\cite{farha2019ms} by using acausal mode with kernel size at 3. Furthermore, we apply the non-linear activation function ReLU to each dilation output followed by a residual connection between the layer input and the convolution signal. The temporal dilated procedure can be formulated as follows:
\begin{equation}
    \hat{E_l} = \mathrm{ReLU} (W_1 * E_{l-1} + b_{1}) 
\end{equation}
\begin{equation}
    E_l = E_{l-1} + W_2 * \hat{E_l} + b_{2}
\end{equation}
where $E_l$ is the output of $l$-th encoder layer, $*$ is the temporal convolutional operation, {$W_1 \in \mathbf{R}^{f\times f\times 3}$, $W_2 \in \mathbf{R}^{f\times f\times 1}$} represent the weights of a dilated convolution and the weights of a $1\times1$ convolution with $f$ convolutional kernels, respectively. $b_1, b_2 \in \mathbf{R}^f$ are denoted as their corresponding biases. In every dilation layer $l$, the receptive field $R$ grows exponentially to capture the long range temporal pattern, expressed as: $R(l) = 2^{l+1} - 1$. By doing this, the temporal information on different scale is hierarchically aggregated while keeps the ordering of sequence. We also employ a $4\times1$ max-pooling layer behind the encoder dilation block to efficiently reduce the oversegmentation error (see our ablative study results in Table~\ref{table_2}).

Our symmetric decoder has a similar structure with the encoder block, except that the max-pooling operations are replaced with a $1\times4$ upsampling. To get the final prediction, we use a $1\times1$ convolution followed by a softmax activation after the last decoder dilated convolution layer:
\begin{equation}
    Y_t = \mathrm{Softmax}(W * D_{L,t} + b)
\end{equation}
where $Y_t$ is the prediction at time $t$, $D_{L,t}$ is the output from the last decode dilated layer at time $t$, $W \in \mathbf{R}^{f\times c}$ and {$b \in \mathbf{R}^c$, where $c \in [1,C]$} is the surgical gestures classes. Eventually, we use the categorical cross-entropy loss for the classification loss calculation.
\subsection{Joint Frame-to-Frame Relation Learning with Self-Attention}\label{SA}
The TCNs have shown consistent robustness in handling long temporal sequences with using \textit{\textbf{relational features}} among frames. However, current methods ~\cite{ding2017tricornet,lea2017temporal} only consider relations in local neighbors, which could undermine their performance in capturing relational features within a longer period. To obtain the global relationship among frames, it is essential to build frame-to-frame relational features with a non-local manner in addition to our encoder-decoder dilated convolutions. With this insight, we introduce the non-local self-attention module to extract discriminate spatial-temporal features for better prediction. 

\begin{figure}[htb!]
\centering
\includegraphics[width=0.5\textwidth]{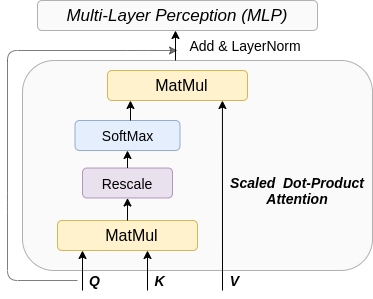}
\caption{Self-attention block}
\label{figure_3}
\end{figure}
Self-attention, or intra-attention refers to an attention mechanism, which attends every position of the input sequence itself and build one-to-one global dependencies. This idea has been widely used in Natural Language Processing (NLP)~\cite{vaswani2017attention}, Object Detection and Segmentation~\cite{wang2018non,hu2018relation}, etc. The key component of self-attention is called \textbf{Scaled Dot-Product Attention}, which is calculated as: 
\begin{equation}
    \mathrm{Attention} (Q,K,V) = \mathrm{Softmax} (\frac{QK^T}{\sqrt{d_k}})V
\end{equation}
where $Q$ is a packed \textit{Query} matrix, $K$ and $V$ stand for \textit{Key-Value} pairs, and $\sqrt{d_k}$ is the feature dimension of queries and keys. The structure of the self-attention is shown in Figure~\ref{figure_3}. In our work, the input \textit{Queries, Keys, and Values} to the self-attention module are the same, that is the output hidden temporal states from the encoder downsampling. The first step is to take the dot product between the query and the key to calculate the similarity. This similarity determines the relevance between all other frames from the input sequence to a certain frame. Then, the dot product is rescaled by $\sqrt{d_k}$ to prevent the exploding gradient and followed by a softmax function to normalize the result. The intention of applying the softmax function here is to give relevant frames more focus and drop irrelevant ones.  Eventually, the attention matrix is multiplied by the value and summed up. There is a residual connection followed by a layer normalization to feed the result to next two fully connected 1-D concolutional layers (see Figure~\ref{figure_3}). In this manner, frame-to-frame global dependencies are constructed.
\section{Evaluation}
\subsection{Experiment Settings}
\textbf{Dataset Description:} We evaluate our approach on an elementary suturing task from JHU-ISI Gesture and Skill Assessment Working Set (JIGSAWS)~\cite{gao2014jhu}, a robotic assisted bench-top model collected using \textit{da Vinci} surgical system. There are 39 videos performed by eight surgeons with three skill levels. Ten different fine-grained surgical gestures for example, \textit{pushing needle through tissue and oriental needle} for suturing task are manually annotated by an experienced surgeon. We follow the standard \textit{leave-one-user-out} (LOUO), a 8-fold cross validation scheme for evaluation. In each fold, we leave one surgeon out for testing to verify if the recognition model works for an unseen subject. For the network input, we use the 128 dimensional spatial-CNN features extracted by~\cite{lea2017temporal} with 10 FPS. Given a video sequence $v \in V$ with length $T$: $v_{1:T} = (v_1,...,v_T)$, our goal is to assign the corresponding gesture label $g \in \mathbf{G}$ to each frame: $g_{1:T}=(g_1,...,g_T)$.

\textbf{Implementation and Training Details:}
The model is implemented based on Pytorch and trained on a single NVIDIA GeForce GTX 1080 graphics card. For the symmetric dilated convolution, we set the layer number to 10 (see the supplementary material for the hyperparameter tuning experiment) and the channel number to 128 with the kernel size 3 followed by a dropout after each layer. In regard to the attention module, the feature dimension of queries and keys is set to 16. The network is trained for 30 epochs with the learning rate at 0.01. In addition, we apply Adam Optimizer such that $\beta_1 = 0.9$, $\beta_2 = 0.98$, and $\epsilon = 10^{-9}$.

\textbf{Evaluation Metrics:} We adopt three evaluation metrics in our experiments: \textbf{frame-wise accuracy}, \textbf{edit score}, and \textbf{segmented F1 score}. 
Frame-wise accuracy is to measure the performance in frame level. However, long gesture segments tend to have more impact than short gesture segments, and the frame-wise accuracy is not sensitive to the oversegmentation error. Therefore, we use the edit score and F1 score to assess the model at segmental level. Edit score is defined as the normalized Levenshtein distance between the prediction and the groundtruth. While F1 score is the harmonic mean of precision and recall with the threshold 10\%, 25\%, and 50\% as defined in~\cite{lea2017temporal}.
\subsection{Comparison with the State-of-The-Arts}
Table~\ref{table_1} compares our symmetric dilation network with other state-of-the-art methods. It can be seen that our model achieves the best performance in all three metrics. Among other approaches, the baseline model \textbf{Bi-LSTM} reaches the relative lower performance than other methods indicating that the traditional RNN-based method is incapable of handing long video sequence. \textbf{Deep Reinforcement Learning (RL)} method trains an intelligent agent with reward mechanism and achieves the high edit 87.96 and F1 score 92.0, but the low frame-wise accuracy at 81.43\%, which shows its inadequacy in capturing the global similarities throughout the frames. The latest \textbf{3D-CNN} method obtains the fair frame-wise accuracy at 84.3\%, but it only obtains 80.0 for the edit score. This reflects that the model based on clip-level is still inefficient in catching long temporal relationship such that it suffers from the oversegmentation error. 

While our model reaches the best frame-wise accuracy at 90.1\% as well as the highest edit and F1 score at 89.9 and 92.5, respectively. It demonstrates that our model is able to capture the long-range temporal information along with the frame-to-frame global dependencies.
\begin{table*}[]
\centering
\caption{Comparsion with the most recent and related works for surgical gesture recognition. Acc., Edit, and F1@{10, 25, 50} stand for the frame-wise accuracy, segmented edit distance, and F1 score, respectively}
\begin{tabular}{llllll}
\hline
\textbf{JIGSAWS (Suturing)} & \textbf{Acc.} & \textbf{Edit} & \textbf{F1@10} & \textbf{F1@25} & \textbf{F1@50} \\
\hline
Bi-LSTM~\cite{singh2016multi}& 77.4& 66.8& 77.8&-&- \\
ED-TCN~\cite{lea2017temporal}& 80.8& 84.7& 89.2&-&-  \\
TricorNet~\cite{ding2017tricornet}& 82.9& 86.8&-&-&-  \\
RL~\cite{liu2018deep}& 81.43& 87.96& 92.0& 90.5 & 82.2\\
3D-CNN~\cite{funke2019using}& 84.3& 80.0& 87.0&-&-\\
Symmetric dilation (w. pooling) + attn & \textbf{90.1} & \textbf{89.9} & \textbf{92.5} & \textbf{92.0} & \textbf{88.2}\\
\hline
\end{tabular}
\label{table_1}
\end{table*}
\section{Discussion}
To further investigate the functionality of each submodule in our method, we conduct ablative studies with five configurations as follows. As our network consists of a symmetric dilation structure with a self-attention kernel in the middle. We decouple it into a head dilation module, a tail dilation module, and the self-attention kernel to explore their joint effects.  
\begin{enumerate}
    \item[(1)] Self-attention module only (baseline)
    \item[(2)] Baseline + head dilated convolution
    \item[(3)] Baseline + tail dilated convolution
    \item[(4)] Baseline + symmetric dilated convolution
    \item[(5)] Baseline + symmetric dilated convolution + pooling
\end{enumerate}

We apply these settings to segment and classify the surgical gestures and measure their \textbf{frame-wise accuracy}, \textbf{edit score}, and \textbf{segmented F1 score} separately. The experiment results are shown in Table~\ref{table_2} (see supplementary material for the visualization of ablative experiments).

\textbf{(1) only:} Self-attention module can achieve promising frame-wise accuracy at 87.8\%, but with very low edit distance (44.0) and F1 scores. It can be concluded that attention module is robust for classification tasks while missing the long temporal information.

\textbf{(1) v.s. (2) and (3):} We put the temporal dilated convolution structure before and after the self-attention module and get the similar results. The results have huge improvement in edit score and F1 score with different threshold, increase around 30\% in each metric. It states that temporal convolution is capable of catching long temporal patterns.

\textbf{(4):} The obvious improvement on the segmental level evaluation shows that the symmetric encoder-decoder dilation structure helps capture the high-level temporal features.

\textbf{(5):} Max-pooling and upsampling further improve the edit distance and F1 score at segmental level such that smooth the prediction and allievate the oversegmentation problem.

Above controlled experiments verify the indispensability of each component for our proposed architecture. From frame-level view, self-attention mechanism is feasible to build non-local dependencies for accurate classification. And from the segmental-level perspective, symmetric dilation with pooling is a viable solution for recognising gestures from long and complicated surgical video data.
\begin{table}[]
\centering
\caption{Ablative experiment results show the effectiveness of each submodel. Acc., Edit, and F1@\{10, 25, 50\}, stand for the frame-wise accuracy, segmented edit distance, and F1 score, respectively}
\begin{tabular}{llllll}
\hline
\textbf{JIGSAWS (Suturing)} & \textbf{Acc.} & \textbf{Edit} & \textbf{F1@10} & \textbf{F1@25} & \textbf{F1@50} \\
\hline
Self-attn only& 87.8 & 44.0 & 54.8 & 53.5 & 49.0 \\
Head dilation + attn& 90.8 & 76.9 & 82.5 & 81.8 & 79.3 \\
Tail dilation + attn& 90.5 & 77.9 & 83.4 & 83.4 & 79.7 \\
Symmetric dilation + attn& 90.7 & 83.7 & 87.7 & 86.9 & 83.6 \\
Symmetric dilation (w. pooling) + attn & 90.1 & \textbf{89.9} & \textbf{92.5} & \textbf{92.0}& \textbf{88.2} \\
\hline
\end{tabular}
\label{table_2}
\end{table}
\section{Conclusion}
In this work, we propose a symmetric dilated convolution network with self-attention module embedded to jointly segment and classify fine-grained surgical gestures from the surgical video sequence. Evaluation of JIGSAW dataset indicates that our model can catch the long-term temporal patterns with the large temporal receptive field, which benefits from the symmetric dilation structure. In addition, a self-attention block is applied to build the frame-to-frame relationship to capture the global dependencies, while the temporal max-pooling and upsampling layer further diminish the oversegmentation error. Our approach outperforms the accuracy of state-of-the-art methods both at the frame level and segmental level. Currently, our network is designed with an off-line manner in acausal mode, and we will explore the possibility of improving and applying it for real-time surgical gesture recognition in the future work.\\\\
\textbf{Acknowledgement} The authors thank Bournemouth University PhD scholarship and Hengdaoruyi Company as well as the the Rabin Ezra Scholarship Trust for partly supported this research.
\bibliographystyle{splncs04}
\bibliography{miccai.bbl}
\clearpage
\textbf{Checklist of the supplementary material:}
\begin{enumerate}
    \item \textbf{Visualization of the ablative experiments.} Figure~\ref{figure_1} is the visualization result of our ablative experiments from the paper, where we decompose our network into five ablated configurations.
    \item \textbf{Influence of the number of dilation layers with visualization.} In Table~\ref{table_1} and Figure~\ref{figure_2}, we set the layer number $l$ to 2, 6, 10, 14 both in encoder and decoder block to explore the impact of the receptive field size. The configuration of 10 symmetric dilation layers achieves the best results.
\end{enumerate}
  
\begin{figure}[ht]
\centering
\includegraphics[width=1.0\textwidth]{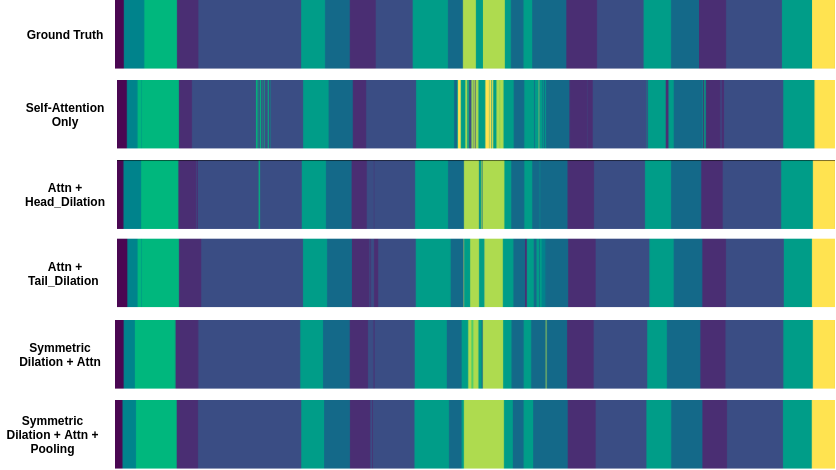}
\caption{Visualization of ablative experiments.(0) ground truth; (1) self-attention module only (baseline); (2) baseline + head dilated convolution; (3) baseline + tail dilated convolution; (4)baseline + symmetric dilated convolution; (5) baseline + symmetric dilated convolution + pooling.}
\label{figure_1}
\end{figure}
\begin{table*}[htb!]
\centering
\caption{Ablative experiment results show the effect of the number of dilation layers (i.e the size of receptive field).  Acc., Edit, and F1@\{10, 25, 50\} stand for the frame-wise accuracy, segmented edit distance, and F1 score, respectively.}
\begin{tabular}{llllll}
\hline
\textbf{JIGSAWS (Suturing)} & \textbf{Acc.} & \textbf{Edit} & \textbf{F1@10} & \textbf{F1@25} & \textbf{F1@50} \\
\hline
2 Layers & 89.6 & 75.0 & 82.2 & 81.3 & 81.3 \\
6 Layers & 90.6 & 88.2 & 91.5 & 91.0 & 87.6 \\
10 Layers & 90.1 & \textbf{89.9} & \textbf{92.5} & \textbf{92.0} & \textbf{88.2} \\
14 Layers & 89.9 & 86.4 & 90.6 & 89.6 & 86.3 \\
\hline
\end{tabular}
\label{table_1}
\end{table*}
\begin{figure}[htb!]
\centering
\includegraphics[width=1.0\textwidth]{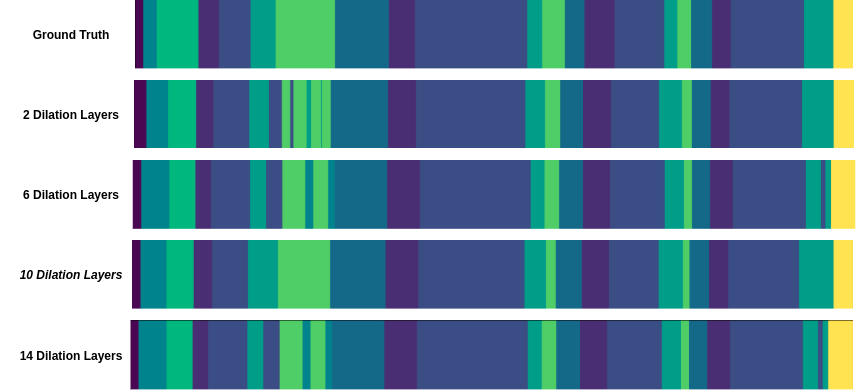}
\caption{Influence of different number of dilation layers. We  set  the layer number $l$ to 2, 6, 10, 14  both in encoder and decoder dilation block.}
\vspace*{3in}
\label{figure_2}
\end{figure}
\end{document}